\documentclass[twoside]{article}

\usepackage{PRIMEarxiv}

\usepackage[utf8]{inputenc} 
\usepackage[T1]{fontenc}    
\usepackage{hyperref}       
\usepackage{url}            
\usepackage{booktabs}       
\usepackage{amsfonts}       
\usepackage{nicefrac}       
\usepackage{microtype}      
\usepackage{lipsum}
\usepackage{fancyhdr}       
\usepackage{graphicx}       
\usepackage{makecell}
\graphicspath{{media/}}     

\newcolumntype{C}[1]{>{\centering\arraybackslash\hspace{0pt}}p{#1}} 
\title{Testing the Segment Anything Model (SAM) on radiology data}

\author{
  José Guilherme de Almeida$^*$ \\
  Computational Clinical Imaging Group \\
  Champalimaud Foundation \\
  Lisbon \\
  Portugal \\
  \texttt{jose.almeida@research.fchampalimaud.pt} \\
   \And
  Nuno Rodrigues\thanks{Equal contribution} \\
  Computational Clinical Imaging Group \\
  Champalimaud Foundation \\
  LASIGE \\
  Faculty of Sciences, University of Lisbon \\
  Lisbon \\
  Portugal \\
  \texttt{nmrodrigues@fc.ul.pt} \\
   \And
  Sara Silva$^\dag$ \\
  LASIGE \\
  Faculty of Sciences, University of Lisbon \\
  Lisbon \\
  Portugal \\
  \texttt{sara@fc.ul.pt} \\
   \And
  Nickolas Papanikolaou\thanks{Shared senior authorship} \\
  Computational Clinical Imaging Group \\
  Champalimaud Foundation \\
  Lisbon \\
  Portugal \\
  \texttt{nickolas.papanikolaou@research.fchampalimaud.org} \\
}

\begin{document}
\maketitle

\begin{abstract}
Deep learning models trained with large amounts of data have become a recent and effective approach to predictive problem solving --- these have become known as "foundation models" as they can be used as fundamental tools for other applications. While the paramount examples of image classification (earlier) and large language models (more recently) led the way, the Segment Anything Model (SAM) was recently proposed and stands as the first foundation model for image segmentation, trained on over 10 million images and with recourse to over 1 billion masks. However, the question remains --- what are the limits of this foundation? Given that magnetic resonance imaging (MRI) stands as an important method of diagnosis, we sought to understand whether SAM could be used for a few tasks of zero-shot segmentation using MRI data. Particularly, we wanted to know if selecting masks from the pool of SAM predictions could lead to good segmentations.

Here, we provide a critical assessment of the performance of SAM on magnetic resonance imaging data. We show that, while acceptable in a very limited set of cases, the overall trend implies that these models are insufficient for MRI segmentation across the whole volume, but can provide good segmentations in a few, specific slices. More importantly, we note that while foundation models trained on natural images are set to become key aspects of predictive modelling, they may prove ineffective when used on other imaging modalities.
\end{abstract}


\section{Introduction}
\label{sec:introduction}

The number of parameters in modern deep learning (DL) models has consistently increased in scale over the past years, with some models now totalling over one trillion parameters \cite{shen2023efficient}. This dramatic increase in scale has led to models capable of solving more complex tasks --- GPT-4, one of the more popular large language models (LLM), has shown remarkable abilities at solving a wide range of exams typically requiring human expertise \cite{openai2023gpt4} (there are of course caveats to these assessments --- details on training protocol and data were made purposefully opaque by OpenAI \cite{openai2023gpt4} and data leakage is a possibility in these exams, which are not realistic ways of evaluating the potential applications of LLM, as noted by Naranayan and Kappor \cite{Sayash2023-nv}), while vision models such as CLIP were notorious for generalising remarkably well to new image classification tasks with no or little finetuning (it should be noted here that CLIP showed poor performance in real-word tasks) \cite{radford2021learning}.

This ability that these large DL models --- otherwise known as foundation models (FM) \cite{Etchemendy_2022-fy} --- display of easily generalising to new tasks with little-to-no adaptation is known as few- or zero-shot learning, respectively. While in few-short learning a small set of examples is provided to the model, in zero-shot learning one expects good performance with no previous examples \cite{Ren2023-sx}. The Segment Anything Model (SAM), first introduced in 2023, is the first foundation segmentation model to date. According to its authors, "its zero-shot performance is impressive" \cite{kirillov2023segment}, making it the ideal candidate for downstream zero-shot segmentation tasks. Particularly, SAM has been shown to perform exceedingly well with different forms of zero-shot learning - not only is it capable of segmenting objects in images with no additional information, one can give prompts (either text or specific points belonging to objects in the image) to guide the segmentation mask prediction.

Radiology is an area of medicine that makes use of medical imaging techniques to address several types of clinical scenarios, such as diagnosis and treatment prediction. Medical imaging techniques such as Magnetic Resonance Imaging (MRI) and Computed Tomography (CT) scans are among the most commonly used due to their high imaging quality and wide array of modalities. There have been several instances of zero-shot learning methods applied to different scenarios of medical imaging, such as x-ray diagnosis~\cite{Paul2021zsxray}, positron image denoising~\cite{Zhu2023xr}, MRI reconstruction~\cite{yaman2022zeroshot} and multi-modality segmentation~\cite{9627926}.

In this work, we assessed the performance of SAM on segmenting some relevant anatomical structures in MRI data. We show that SAM performs poorly, not being able to offer valid segmentation proposals in most cases.

\section{Context}
\label{sec:context}

\subsection{Foundation models}

Foundation models is the generic term used to describe models which can be used on a number of downstream tasks, acting as a common basis \cite{bommasani2022opportunities}. In a way, this term is self-aggrandising \cite{Marcus_2021-kr}, but it captures in broad strokes the purpose of their development --- acting as a large and ultimately robust model on top of which other applications can be built. While earlier foundation models for image-based tasks were trained using classification objectives \cite{Ridnik_2021,bommasani2022opportunities}, more recent image foundation models focus on hybrid tasks or self-supervised learning \cite{wang2022omnivlone, yu2022coca, yuan2021florence, xie2021selfsupervised}. To highlight their potential popularity in clinical predictive modelling, we note that, recently, Moor \textit{et al.} proposed in Nature the development of foundation models for "generalist medical artificial intelligence" \cite{Moor2023-fa}; in other words, models capable of generating medical insight --- explanations, recommendations or, as we explore here, image annotations.

\subsection{Zero-shot segmentation}

Zero-shot segmentation involves the generation of masks for new, previously unseen classes \cite{bucher2019zeroshot}, with several different methods arising in recent years. Bucher and others use a text model to embed classes which are mapped into a feature space common to both pixel embeddings and text, allowing them to map new words (classes) to pixel embedding space \cite{bucher2019zeroshot}.
Björn et. al~\cite{9665941} use a generative model to learn the distribution of the unseen classes, which included a novel decoder architecture that allowed for the generation of class-specific features. This model would then use descriptions of the targets to generate features that would be used by a classifier to perform the segmentation.
Luddecke et. al~\cite{Luddecke_2022_CVPR} use a dual-encoder CLIP-based~\cite{pmlr-v139-radford21a} architecture to handle both image prompts as well as supporting text prompts that share a common decoder used for the segmentation.
Bian et. al~\cite{9627926} propose a two-stage approach for multi-modality medical image segmentation, where during the first stage, a fully supervised CNN is trained to extract relation prototypes from a given modality, which are then used to train the zero-shot model for additional modalities.

\subsection{SAM in the medical domain}

He et. al~\cite{he2023computervision} made a large and thorough comparison between the performance of SAM, and that of regular biomedical segmentation models, such as UNet, on 12 medical imaging problems. This set included six different medical imaging modalities and ten distinct target organs. The authors conclude that, as expected, SAM is not competitive with other segmentation models tailored for both 2D and 3D medical images, achieving significantly worse performance in all tasks.
Zhang et. al~\cite{zhang2023customized} propose finetuning the image-encoder, prompt-decoder and mask-decoder part of SAM for medical segmentation tasks. The authors use LoRA~\cite{hu2021lora} finetuning strategy and evaluate it on a dataset of 30 abdominal CT scans. The results show that that their approach can reach competitive results when compared to regular biomedical segmentation models on this small dataset. Similarly, Ma et. al~\cite{ma2023segment} also propose finetunning SAM, albeit in their case, only the mask-decoder e finetunned and the details  of it are not included in the paper.

\subsection{Annotation of 3D radiology data}

 Predictive modeling methods for radiology volumes (three dimensional images) are a relatively popular research topic \cite{Hosny2018-pd}. However, the field has been hindered by lack of human-annotated data --- on the one hand, human annotation is labour intensive and requires highly specialised knowledge; on the other, it is important and necessary to ensure that patient privacy is preserved, so data sharing is currently only done under strict rules \cite{Willemink2020-rs}; to further complicate this issue, even when data can be shared, biomedical researchers are often not trained on how to share their data as highlighted in a 2015 survey \cite{Federer2015-hu}. While some efforts are undergoing to see this lack of data ameliorated \cite{Sunoqrot2022-zm,Cushnan2021-zx}, the datasets used in training DL models for radiology seldom go beyond hundreds or a few thousand examples \cite{Menze2015-kv,Saha2022-ki}. It should be noted that this is not the case for two-dimensional images. For instance, MedMNIST v2, a large dataset with several different biomedical imaging modalities, makes this contrast apparent --- while some of the 2D radiology datasets contain over 100,000 images, the largest 3D radiology dataset is obtained from 1,018 exams \cite{Yang2023-hy}. These reasons combined led to the development of a large number of models \cite{Liu2021-fz} whose generalisation capabilities are, ultimately, hard to assess. Taking all of this into consideration, the appeal of foundation models with zero-shot learning --- models capable of, in theory, helping radiologists and other experts segment 3D radiology data --- becomes apparent.

\section{Methods}

\subsection{Data}
\label{sec:methods-data}

We used primarily two data sources: the Medical Segmentation Decathlon \cite{Antonelli2022-yk} and ProstateX \cite{Armato2018-jv}. The Medical Segmentation Decathlon contains radiology (MRI or CT) data for 10 different tasks, but we focus on the 8 tasks featuring manual monomodal data and prostate, for which we use the T2 sequence alone (in effect this only excludes the first task of brain tumour segmentation). For ProstateX, a multimodal prostate MRI dataset, we focused on T2-weighted images as these have been shown to yield the best performance \cite{Zabihollahy2019-ep}. Additional information regarding the data can be seen in Table~\ref{tab:data}

\begin{table}[]
\centering
\setlength{\tabcolsep}{10pt} 
\renewcommand{\arraystretch}{1.5} 
\caption{Specifications regarding the datasets used.}
\begin{tabular}{lcccc}
\multicolumn{1}{c}{Dataset} & Source & Modality & \# Samples & Task                               \\ \hline
ProstateX                   & \cite{Armato2018-jv}     & MRI      & 181        & Gland                              \\
Prostate                    & \cite{Antonelli2022-yk}     & MRI      & 32         & \makecell{Peripheral and \\ transition zone}     \\
Heart                       & \cite{Antonelli2022-yk}     & MRI      & 20         & Left atrium                        \\
Spleen                      & \cite{Antonelli2022-yk}     & CT       & 41         & Spleen                             \\
Colon                       & \cite{Antonelli2022-yk}     & CT       & 126        & Primary colon cancer               \\
Hippocampus                 & \cite{Antonelli2022-yk}     & MRI      & 2660       & \makecell{Anterior and \\ posterior hippocampus}
\end{tabular}
\label{tab:data}
\end{table}

\subsection{SAM inference}
\label{sec:methods-inference}

While SAM is a two-dimensional model trained on three channel (RGB) data, MRI data is three-dimensional data with a single channel. To make MRI data compatible with SAM, we first normalize each image by scaling all pixel values to be between 0 and 255 ($p'_{ij} = 255 * \frac{p_{ij} - \min{P}}{\max{P} - \min{P}}$, where $p_{ij}$ and $p'_{ij}$ are a pixel value before and after normalization, respectively, and $P$ is the set of pixel values in a given image), and convert pixel values to unsigned 8-bit integers. Finally, we stack the same modality three times on the channel dimension. This ensures that data quantization is compatible between our data and the data that was used to train SAM. Then, we predict segmentation masks on individual slices. To identify the slice dimension, we assume that the dimension with the smallest size is the slice dimension. If the MRI data is a cube, we select the last dimension as the slice dimension. All experiments were ran using and building on the code available on the Facebook Research GitHub \footnote{https://github.com/facebookresearch/segment-anything} with the weights for the ViT-H.

For inference, we use two distinct zero-shot segmentation methods with SAM, each of which can produce between 0 and several masks (we outline the mask selection process in \ref{sec:methods-selection}): 

\begin{itemize}
    \item \textbf{Standard SAM} --- feeding the image to the model using the \texttt{SamAutomaticMaskGenerator.generate} function;
    \item \textbf{Seeded SAM} --- feeding the image to the model and using an altered version of \texttt{SamAutomaticMaskGenerator}, \texttt{SAMMaskGeneratorWithPoints} (a link to the code can be found in \ref{sec:methods-code}), which allows use to run the \texttt{SamAutomaticMaskGenerator} with a set of points (or seeds). To generate each one of these seeds, we calculate a grid of 16 equally spaced points along the object. We tested seeded SAM only on ProstateX and note that this is similar to one of the zero-shot case studies provided by the authors of SAM --- in section 7.1 of the SAM paper, the authors show that SAM is capable of detecting new objects with a single seed \cite{kirillov2023segment}.
\end{itemize}
 
\subsection{Selection of segments of interest}
\label{sec:methods-selection}

Given that SAM can produce a relatively large number of segment predictions for each image (slice), we focus here on a relatively specific use case --- we wish to run the inference a single time with either standard or seeded SAM. Upon this, we identify which predicted segments belong to the anatomical structure of inference using four heuristics:

\begin{itemize}
    \item \textbf{All SAM masks} --- any segment with a non-zero intersection over union (IoU) with the ground truth is selected. The resulting mask is then the union of these segments;
    \item \textbf{Best SAM masks} --- the segment with the best IoU with the ground truth is selected;
    \item \textbf{SAM masks IoU>0.25} --- segments with $\mathrm{IoU} > 25\%$ with the ground truth are selected;
    \item \textbf{SAM masks IoU>0.5} --- segments with $\mathrm{IoU} > 50\%$ with the ground truth are selected.
\end{itemize}

We note here that this is a very generous evaluation of a segmentation model --- typically, no "oracular" information on the ground truth is provided. However, the use case we propose here mimics that of an end user running SAM once and selecting the best segment(s). We call the final prediction obtained with any of the four heuristics outlined above the "segment of interest".

\subsection{Performance evaluation}

Having used either of the four heuristics outlined above, we focus on the Dice score ($\mathrm{Dice} = \frac{2 \times \mathrm{TP}}{2 \times \mathrm{TP} + \mathrm{FP} + \mathrm{FN}}$, where $\mathrm{TP}$, $\mathrm{FP}$ and $\mathrm{FN}$ are the true positives, false positives and false negatives, respectively). To assess whether the model performs well on specific slices, we also calculate the Dice score for each volume considering only slices with at least one valid predicted segment (we henceforth refer to this as $\mathrm{Dice_{pos}}$). Finally, we also calculate the ratio between the number of ground truth pixels in slices where at least one segment of interest was found and the total number of ground truth pixels. We term this last quantity the "fraction of detected object". All metrics are calculated individually for each sample and the metrics distributions for each zone and dataset are analysed.

\subsection{Code availability}
\label{sec:methods-code}

The code used to run these experiments is available on GitHub \footnote{https://github.com/josegcpa/sam-biomed-img}. The specific implementation of \texttt{SamAutomaticMaskGenerator} with seed points is present in \texttt{med\_img\_sam/sam\_wrappers.py}. We note that, while it is possible to use the native \texttt{SamPredictor} for this, we wanted to maintain the standard format produced by \texttt{SamAutomaticMaskGenerator}; for this reason, introducing minor changes to this function was the easiest way of implementing this.

\section{Results}

\subsection{SAM underperforms on radiology data}

\begin{figure}
    \centering
    \includegraphics[width=\linewidth]{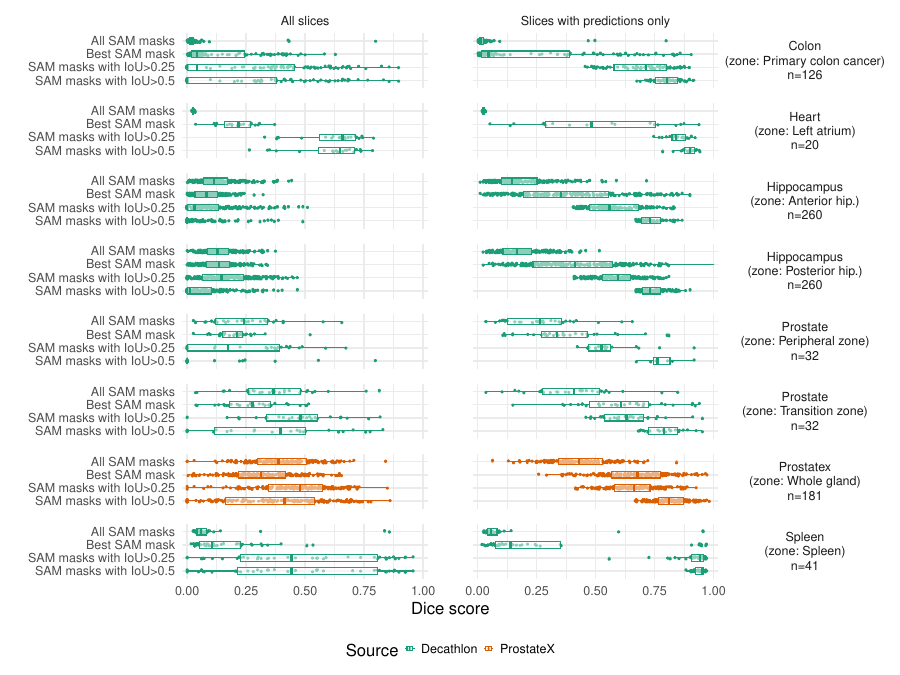}
    \vspace*{-10pt}\caption{Intersection over union for the Segment Anything Model (SAM) when considering all slices in a volume (left) and when considering only slices with predictions. Each vertical facet corresponds to an anatomical zone and colours represent the source of the data (between ProstateX and the Medical Segmentation Decathlon datasets).}
    \label{fig:performance}
\end{figure}

\begin{figure}
\centering
    \includegraphics[width=0.4\linewidth]{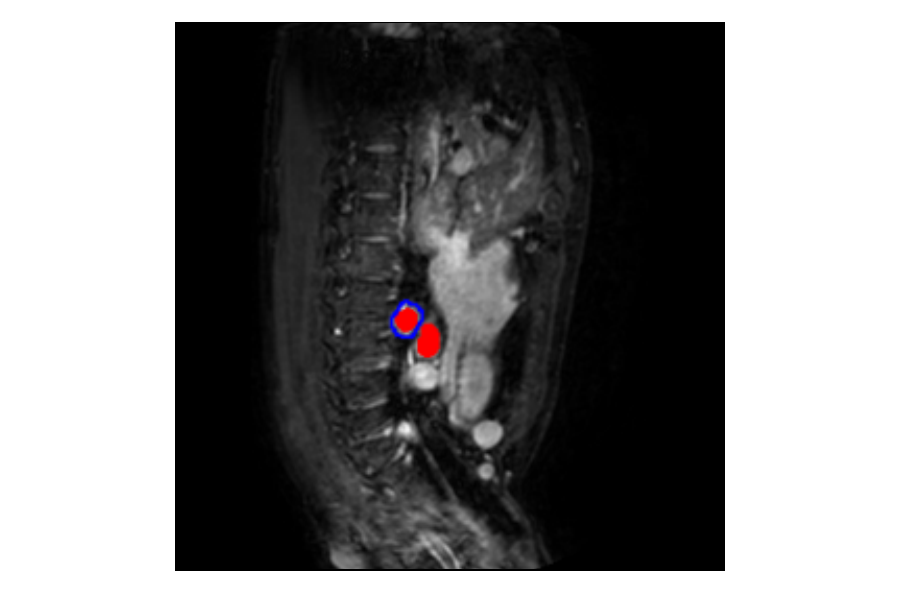}\hspace{-7em}
    \includegraphics[width=0.4\linewidth]{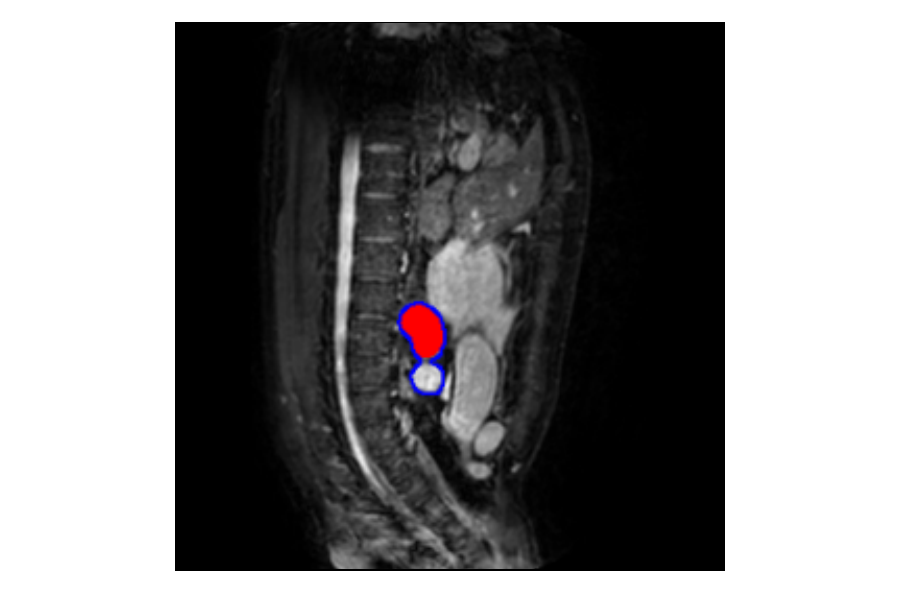}\hspace{-7em}%
    \includegraphics[width=0.4\linewidth]{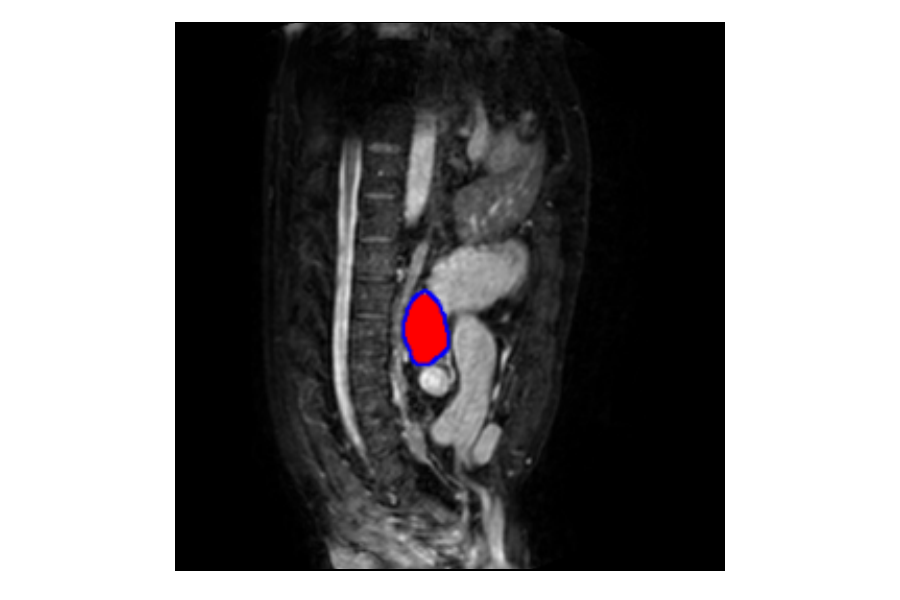}\\\vspace{-1em}
    
    \includegraphics[width=0.4\linewidth]{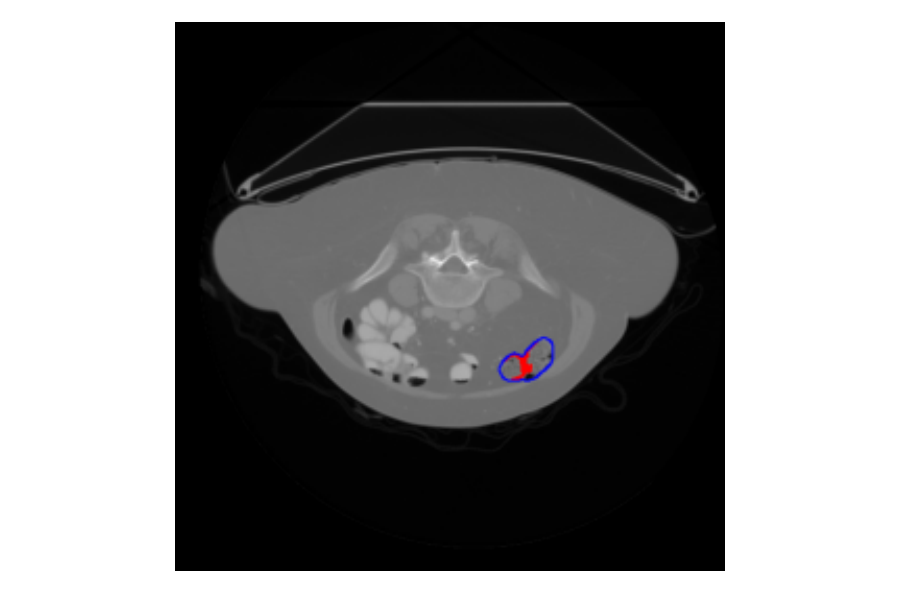}\hspace{-7em}
    \includegraphics[width=0.4\linewidth]{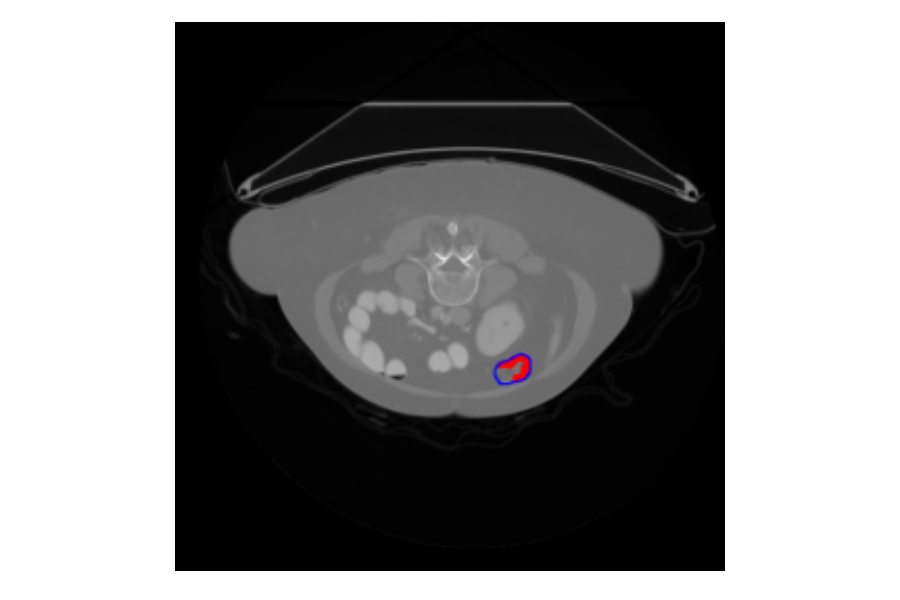}\hspace{-7em}%
    \includegraphics[width=0.4\linewidth]{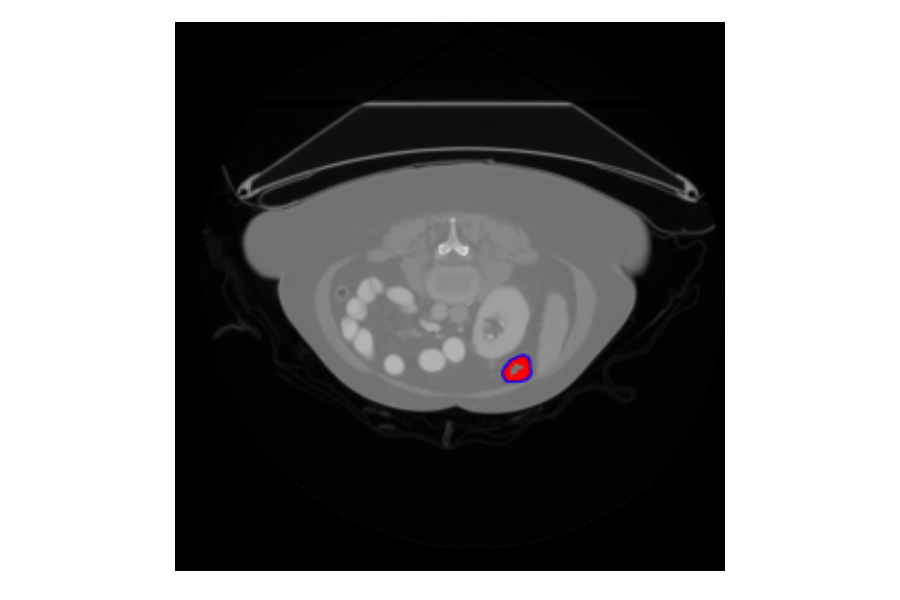}\\\vspace{-1em}

    \includegraphics[width=0.4\linewidth]{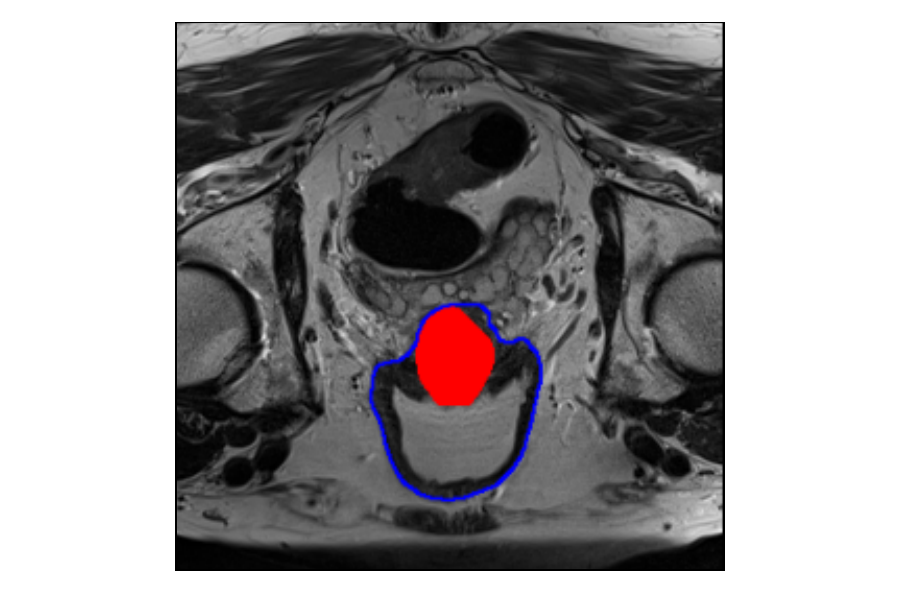}\hspace{-7em}
    \includegraphics[width=0.4\linewidth]{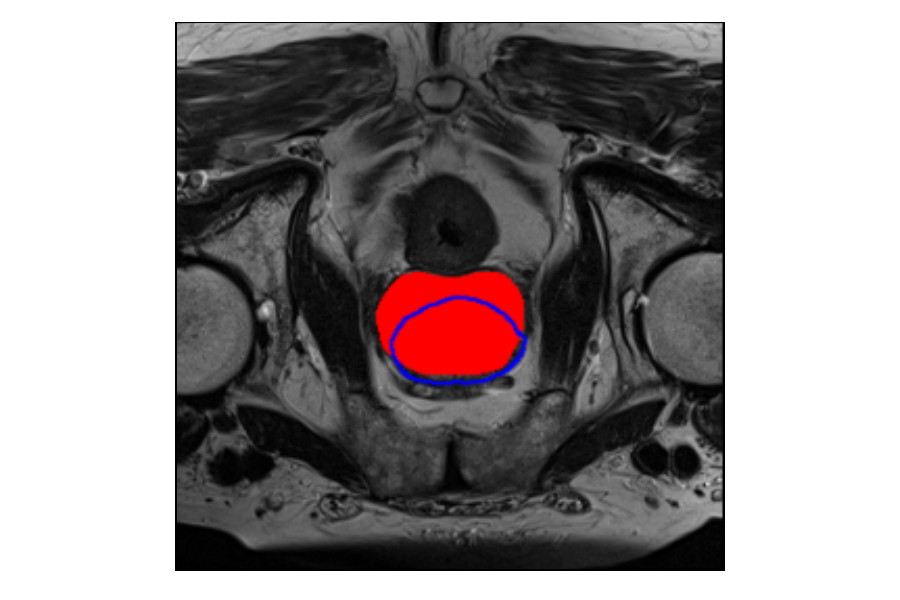}\hspace{-7em}%
    \includegraphics[width=0.4\linewidth]{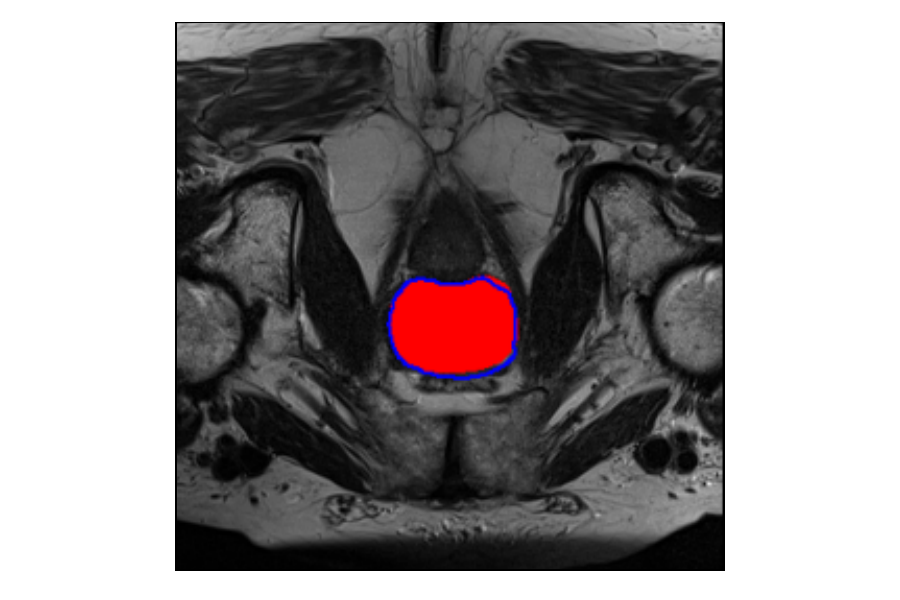}\\\vspace{-1em}

    \includegraphics[width=0.4\linewidth]{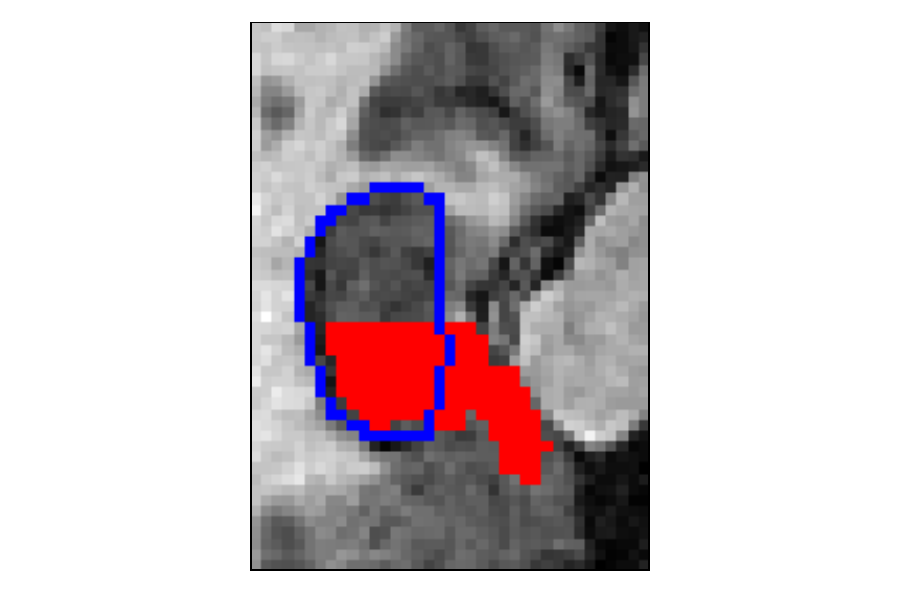}\hspace{-7em}
    \includegraphics[width=0.4\linewidth]{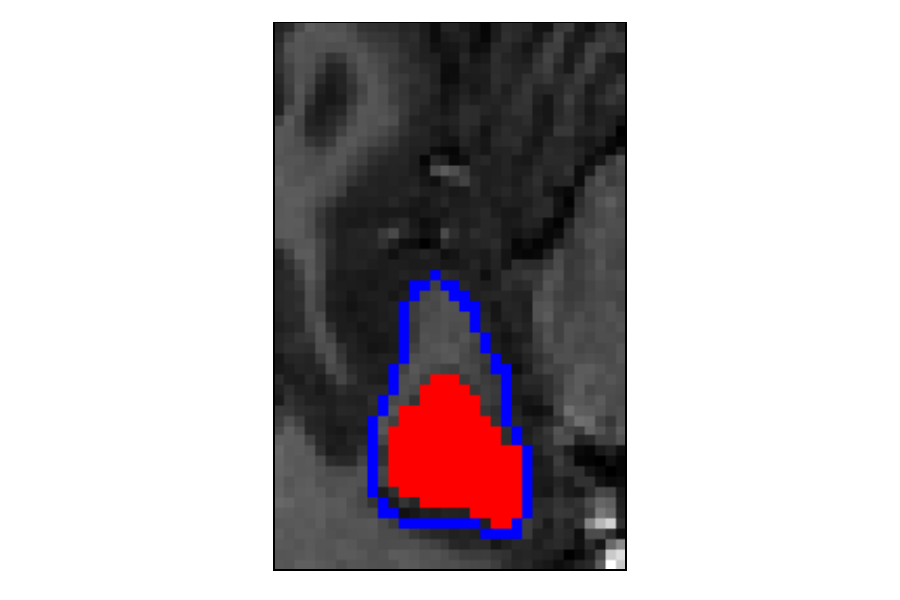}\hspace{-7em}%
    \includegraphics[width=0.4\linewidth]{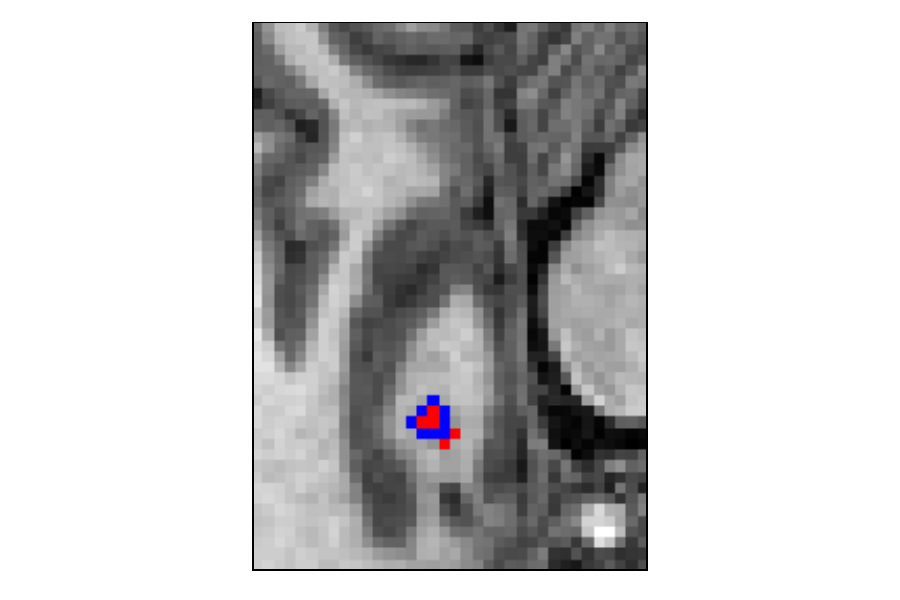}\\\vspace{-1em}

    \includegraphics[width=0.4\linewidth]{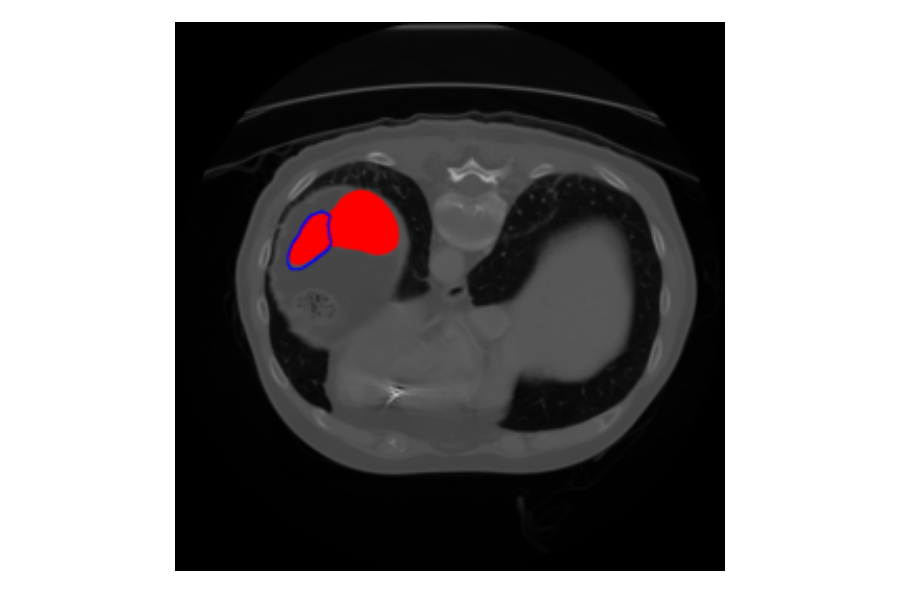}\hspace{-7em}
    \includegraphics[width=0.4\linewidth]{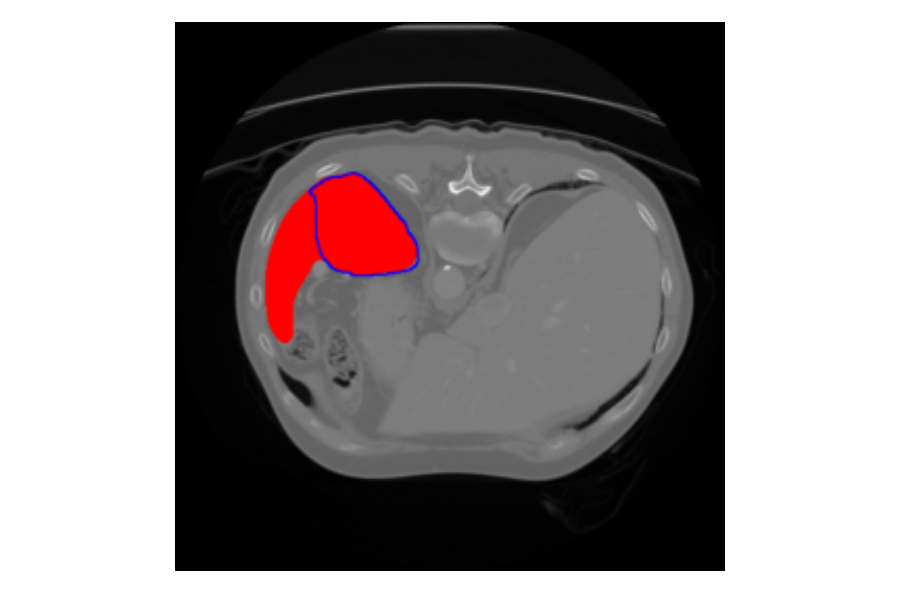}\hspace{-7em}%
    \includegraphics[width=0.4\linewidth]{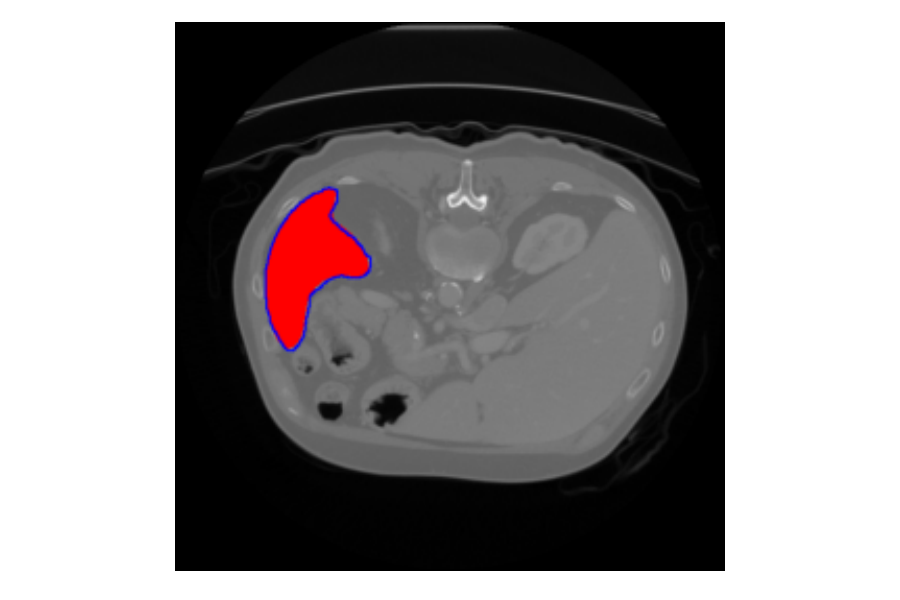}\\
    \begin{tabular}{C{4cm}C{4cm}C{4cm}}
    \textbf{$\approx$25\%} & \textbf{$\approx$50\%} & \textbf{Best}
    \end{tabular}

    \vspace*{-5pt}\caption{Examples of SAM’s segmentations of medical images sorted by IoU. The blue outlines represent the ground truth, while the red binary masks represent the outputted segmentation by SAM.}
    \label{fig:test}
\end{figure}

We applied SAM to 5 different radiology segmentation datasets and calculated the IoU with the ground truth for each individual sample. In general, we observe that SAM does not perform particularly well when segmenting any of the relevant anatomical structures across datasets (Figures~\ref{fig:performance}, left, and \ref{fig:test}). Indeed, the best performance is observed for the segmentation of the left atrium in heart MRI data ($\mathrm{Dice} = 65\%$ when considering masks with > 0.25 IoU) --- in other words, if one accepts segments with a relatively small overlap, it is possible to get a perhaps acceptable segmentation with SAM. Considering the best performance for each dataset, SAM performs the worst when detecting colon cancer in colon CT ($\mathrm{Dice} = 4.2\%$). We do however admit that this is a particularly challenging segmentation task, with the best performing model of the Medical Segmentation Decathlon achieving a Dice score of 56\%, the lowest in the challenge \cite{Antonelli2022-yk}.

\subsection{SAM --- potentially useful in limited cases?}

To understand whether SAM could be helpful in a more limited setting --- providing initial segmentation predictions for a few slices --- we calculate the Dice score considering only the slices where any given prediction is available. We show in Figure~\ref{fig:performance} (right) that this may be a more acceptable, albeit limited, use case for SAM. Indeed, when considering masks with a relatively high overlap with the ground truth ($\mathrm{IoU} > 0.5$), SAM achieves relatively good median performances ($\mathrm{Dice_{pos}} = 95\%$ and $\mathrm{Dice_{pos}} = 90\%$ for spleen and left atrium segmentation, respectively). However, we note that this is quite a unique setting and use case --- a researcher using SAM to segment only a few slices, keeping only those where SAM is capable of producing an acceptable segmentation mask. In other words, a potential user of SAM in radiology would have to be prepared to consider only the select segmentation masks in the few cases where SAM was capable of predicting a minimally overlapping mask.

\begin{figure}
    \centering
    \includegraphics[width=\linewidth]{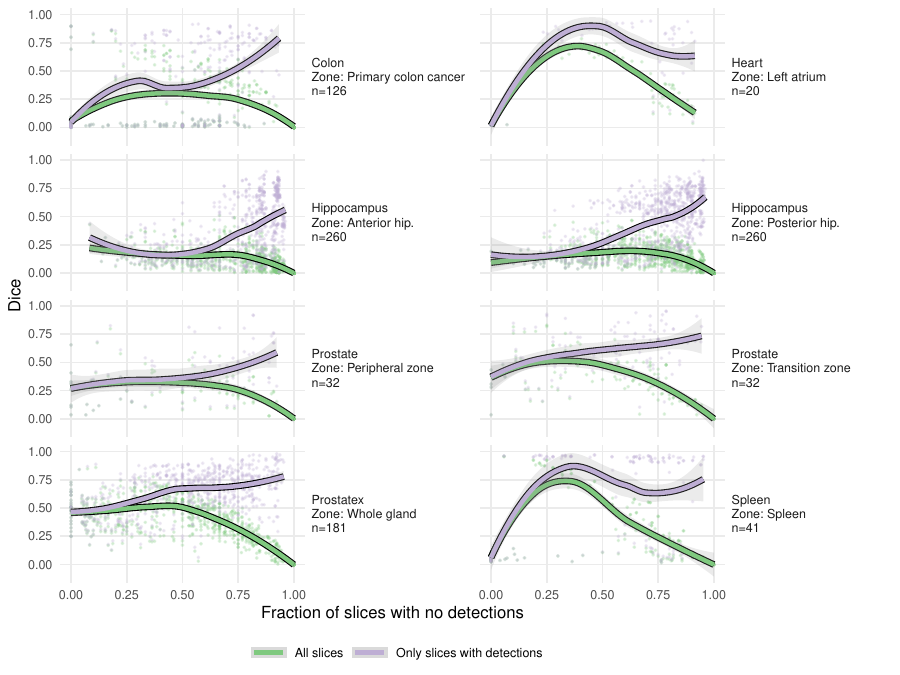}
    \vspace*{-10pt}\caption{Dice scores for all predictions and its association with the number of slices with no segment detections. Each panel corresponds to an anatomical zone for a given dataset and colours correspond to Dice scores calculated using all slices (green) or only slices with detections (purple). The linear fit was achieved with locally estimated scatterplot smoothing (otherwise known as "loess"). Each point represents one volume.}
    \label{fig:lines}
\end{figure}

\begin{figure}
    \centering
    \includegraphics{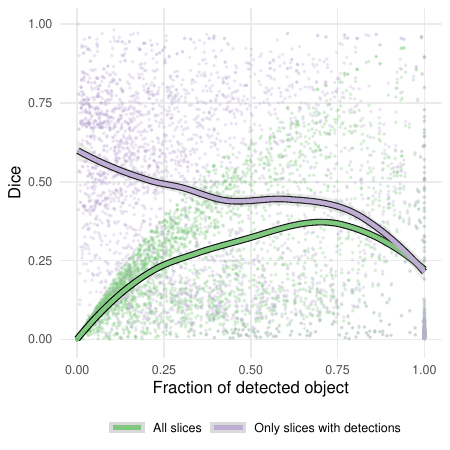}
    \vspace*{-10pt}\caption{Dice scores for all predictions and its association with the fraction of the object which was detected (dividing the number of pixels in the slices with the detected object by the total object size). The linear fit was achieved with locally estimated scatterplot smoothing (otherwise known as "loess"). Each point represents one volume (here, all 660 volumes are plotted).}
    \label{fig:lines-all}
\end{figure}

To better understand this, we also study how this "selective performance" is influenced by the fraction of slices where no overlapping segmentation was predicted by SAM. In Figure~\ref{fig:lines} we highlight this, showing that the average $\mathrm{Dice_{pos}}$ increases with the fraction of slices with no detection in most datasets (an exception is observable --- left atrium segmentation, which may indicate that the left atrium is easier to predict in specific slices). In Figure~\ref{fig:lines-all}, we further highlight this through the inspection of the association between $\mathrm{Dice_{pos}}$ and the fraction of detected object (here calculated simply as the fraction of the object contained in slices where SAM produced a segmentation with $\mathrm{IoU} > 0$) --- as the fraction of the detected object increases, $\mathrm{Dice_{pos}}$ has a tendency to decrease. 

\subsection{Seeds do not lead to improved performance}

\begin{figure}
    \centering
    \includegraphics[width=\linewidth]{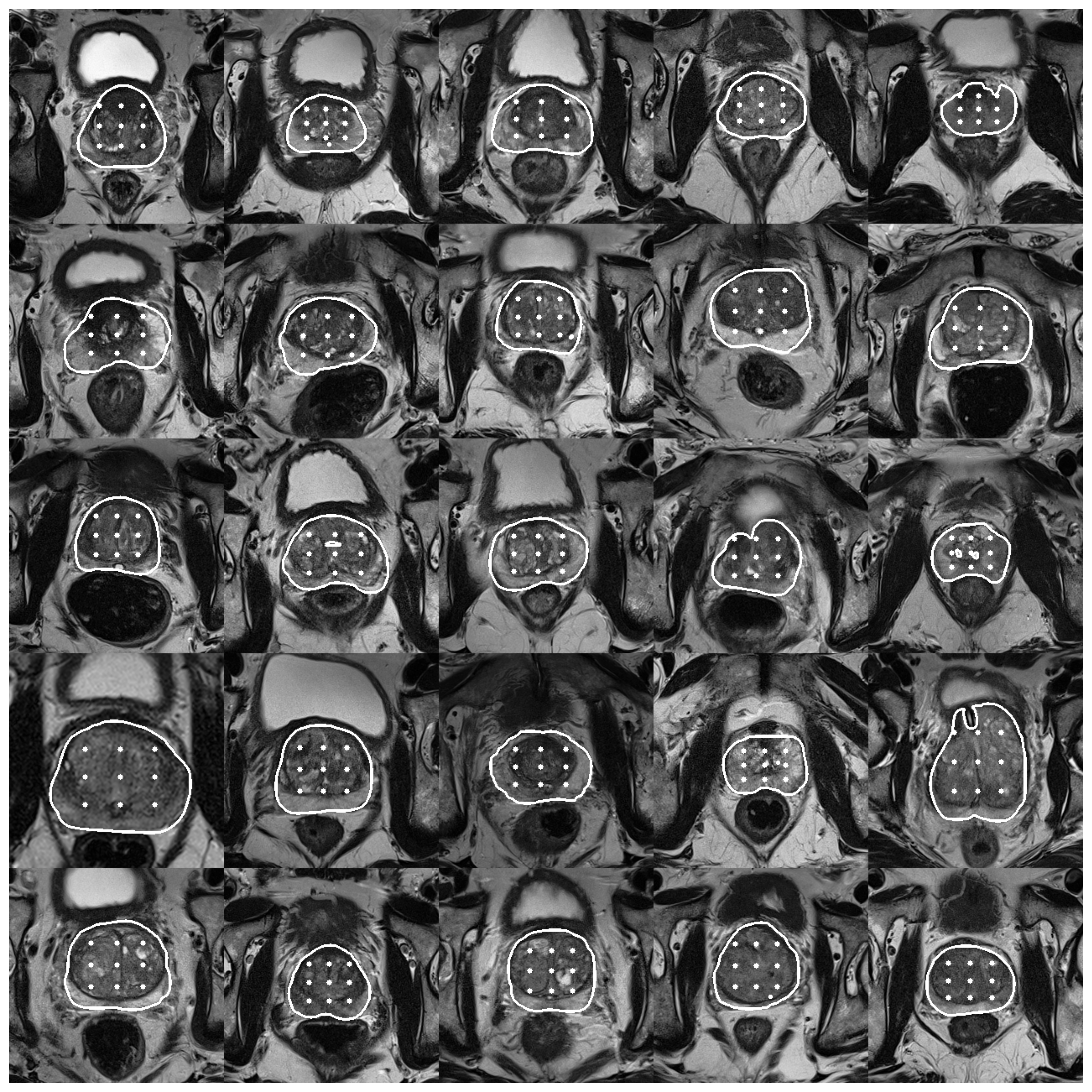}
    \vspace*{-10pt}\caption{Visualization of the $3 \times 3$ grid of seeds (point prompts) used as input to the SAM model using the ProstateX dataset. Points represent seeds and the white outline represents the contour of the prostate segmentation ground truth.}
    \label{fig:seeded_viz}
\end{figure}

\begin{figure}
    \centering
    \includegraphics{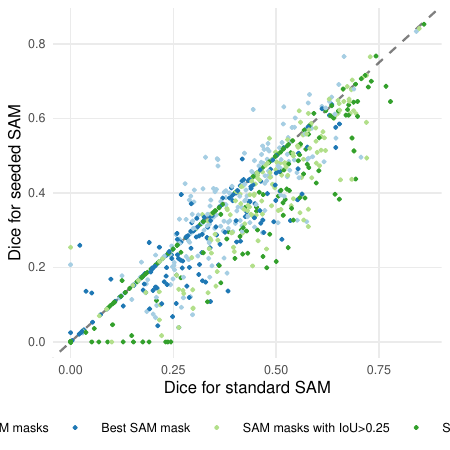}
    \vspace*{-10pt}\caption{Comparison of the Dice scores obtained using standard SAM and seeded SAM. Each point represents a study in ProstateX (n=181) and colours correspond to different segment selection heuristics.}
    \label{fig:seeded}
\end{figure}

To test whether seeds (point prompts for SAM) can ameliorate the poor performance observed thus far, we conducted a simple experiment: using the ProstateX dataset, we take the available whole gland masks and calculate a $3 \times 3$ grid of equally spaced points spanning the whole prostate (Figure~\ref{fig:seeded_viz}). Using these points, we prompt SAM to obtain prostate gland masks. We observe that this fails to improve Dice scores (Figure~\ref{fig:seeded}). In fact, controlling for segment selection heuristic, we observe an unexpected drop of 4\% Dice when using seeded SAM (Table \ref{tab:linear-seeded}). This suggests that, for radiology, SAM benefits more from having a more aggressive seeding mechanism (by default SAM uses a $32 \times 32$ grid of equally spaced coordinates as seeds) than from having more targeted seeds. 

\begin{table}[hb!]
\caption{Effect of seeded SAM and different heuristics on Dice scores. To infer these coefficients, a linear model with Dice scores as the dependent variable and categorical factors for SAM type (standard SAM vs. seeded SAM) and heuristics (all masks vs. best mask vs. SAM masks with IoU>0.25 vs. SAM masks with IoU>0.5).}
\centering
\begin{tabular}{rrrrr}
    & Estimate & Std. Error & $t$-value & $p$-value \\ 
    \hline
    Intercept (standard SAM + all masks) & $0.41$ & $0.010$ & $39.25$ & $7.09 \times 10^{-230}$ \\ 
    Seeded SAM & $-0.04$ & $0.009$ & $-4.75$ & $2.19 \times 10^{-6}$ \\ 
    Best SAM mask & $-0.09$ & $0.013$ & $-6.71$ & $2.74 \times 10^{-11}$ \\ 
    SAM masks with $\mathrm{IoU}>0.25$ & $0.04$ & $0.013$ & $2.78$ & $5.44 \times 10^{-3}$ \\ 
    SAM masks with $\mathrm{IoU}>0.5$ & $-0.05$ & $0.013$ & $-3.82$ & $1.38 \times 10^{-4}$ \\ 
    \hline
    \label{tab:linear-seeded}
\end{tabular}
\end{table}

\section{Limitations}

It is important to highlight how this work is, to some extent, an unfair comparison --- SAM was trained on natural images (i.e. photographic images of the real world), making this a challenging task \textit{a priori}: individual voxel values in radiology data are encoded as values which depend on the particular image acquisition process, and the relationships between different voxels is unlikely to be similar to that found in photographic images. Additionally, the contrast and relative intensities in radiology images are oftentimes considerably different to that found in natural images. 

For these reasons, we note that this work is not meant to be a damning critique of the capabilities of SAM; instead, we hope this work grounds the expectations of radiologists and computational researchers alike when applying SAM to radiology. Other recently developed adaptations of SAM, such as those by Ma and Wang \cite{ma2023segment} or by Zhang and Liu \cite{zhang2023customized} have shown to be of superior quality in this domain.

\section{Conclusion}

We first considered this assessment as a curiosity --- we not only hoped to use SAM, but also wanted to better understand how SAM, a foundation model, could be used to segment anatomical structures in radiology data. However, the results here shown highlight what other authors noted when applying SAM to digital pathology: the results can be relatively dissatisfying \cite{deng2023segment}. Finally, we wish to note that the terminology and discussion surrounding "foundation models" often lead to overarching claims about their capabilities; in a domain with imaging modalities as varied as computer vision, it is important to carefully assess the capabilities of each model across different domains (by this we do not wish to imply that every research paper on computer vision should test on every imaging modality; just that these assessments constitute an important part of research, particularly on foundation models).

Models such as SAM show a wide range of impressive applications when working with natural images, but it is important to consider how exactly they can be applied outside of this domain. We show here that their application in radiology is limited --- indeed, while an end user can perhaps obtain a set of acceptable predictions for their anatomical region of interest, the model oftentimes fails to pick up any relevant segments in the image, rendering its application in automated segmentation elusive for the time being. Considering what has been presented thus far, we recommend the critical and careful use of SAM in radiology, highlighting that while it may not be able to segment anything under all circumstances, it may be able to offer acceptable predictions in some instances. Additionally and as noted earlier, SAM-based foundational models by others have shown potential in this field \cite{ma2023segment,zhang2023customized}. However, it should be highlighted that performance is still lacking when compared with domain- and anatomy-specific models.

\section*{Acknowledgments}

JGA is supported by the European Union H2020: ProCAncer-I project (EU grant 952159).
This work was partially supported by the FCT, Portugal, through funding of the LASIGE Research Unit (refs.~UIDB/00408/2020 and UIDP/00408/2020). Nuno M.~Rodrigues was supported by PhD Grant 2021/05322/BD

\bibliographystyle{unsrt}  
\bibliography{references}

\begin{thebibliography}{10}

\bibitem{shen2023efficient}
Li~Shen, Yan Sun, Zhiyuan Yu, Liang Ding, Xinmei Tian, and Dacheng Tao.
\newblock On efficient training of large-scale deep learning models: A
  literature review, 2023.

\bibitem{openai2023gpt4}
OpenAI.
\newblock Gpt-4 technical report, 2023.

\bibitem{Sayash2023-nv}
Narayanan Sayash and Kapoor Arvind.
\newblock {GPT-4} and professional benchmarks: the wrong answer to the wrong
  question.
\newblock
  \url{https://aisnakeoil.substack.com/p/gpt-4-and-professional-benchmarks},
  March 2023.
\newblock Accessed: 2023-4-17.

\bibitem{radford2021learning}
Alec Radford, Jong~Wook Kim, Chris Hallacy, Aditya Ramesh, Gabriel Goh,
  Sandhini Agarwal, Girish Sastry, Amanda Askell, Pamela Mishkin, Jack Clark,
  Gretchen Krueger, and Ilya Sutskever.
\newblock Learning transferable visual models from natural language
  supervision, 2021.

\bibitem{Etchemendy_2022-fy}
John Etchemendy.
\newblock Introducing the center for research on foundation models ({CRFM}).
\newblock
  \url{https://hai.stanford.edu/news/introducing-center-research-foundation-models-crfm},
  2023.
\newblock Accessed: 2023-4-17.

\bibitem{Ren2023-sx}
Wenqi Ren, Yang Tang, Qiyu Sun, Chaoqiang Zhao, and Qing-Long Han.
\newblock Visual semantic segmentation based on few/zero-shot learning: An
  overview.
\newblock {\em IEEE/CAA Journal of Automatica Sinica}, pages 1--21, 2023.

\bibitem{kirillov2023segment}
Alexander Kirillov, Eric Mintun, Nikhila Ravi, Hanzi Mao, Chloe Rolland, Laura
  Gustafson, Tete Xiao, Spencer Whitehead, Alexander~C. Berg, Wan-Yen Lo, Piotr
  Dollár, and Ross Girshick.
\newblock Segment anything, 2023.

\bibitem{Paul2021zsxray}
A.~Paul, T.~C. Shen, S.~Lee, N.~Balachandar, Y.~Peng, Z.~Lu, and R.~M. Summers.
\newblock {{G}eneralized {Z}ero-{S}hot {C}hest {X}-{R}ay {D}iagnosis {T}hrough
  {T}rait-{G}uided {M}ulti-{V}iew {S}emantic {E}mbedding {W}ith
  {S}elf-{T}raining}.
\newblock {\em IEEE Trans Med Imaging}, 40(10):2642--2655, Oct 2021.

\bibitem{Zhu2023xr}
Mingwei Zhu, Min Zhao, Min Yao, and Ruipeng Guo.
\newblock A generative adversarial network with ``zero-shot'' learning for
  positron image denoising.
\newblock {\em Scientific Reports}, 13(1):1051, January 2023.

\bibitem{yaman2022zeroshot}
Burhaneddin Yaman, Seyed Amir~Hossein Hosseini, and Mehmet Akcakaya.
\newblock Zero-shot self-supervised learning for {MRI} reconstruction.
\newblock In {\em International Conference on Learning Representations}, 2022.

\bibitem{9627926}
Cheng Bian, Chenglang Yuan, Kai Ma, Shuang Yu, Dong Wei, and Yefeng Zheng.
\newblock Domain adaptation meets zero-shot learning: An annotation-efficient
  approach to multi-modality medical image segmentation.
\newblock {\em IEEE Transactions on Medical Imaging}, 41(5):1043--1056, 2022.

\bibitem{bommasani2022opportunities}
Rishi Bommasani, Drew~A. Hudson, Ehsan Adeli, Russ Altman, Simran Arora, Sydney
  von Arx, Michael~S. Bernstein, Jeannette Bohg, Antoine Bosselut, Emma
  Brunskill, Erik Brynjolfsson, Shyamal Buch, Dallas Card, Rodrigo Castellon,
  Niladri Chatterji, Annie Chen, Kathleen Creel, Jared~Quincy Davis, Dora
  Demszky, Chris Donahue, Moussa Doumbouya, Esin Durmus, Stefano Ermon, John
  Etchemendy, Kawin Ethayarajh, Li~Fei-Fei, Chelsea Finn, Trevor Gale, Lauren
  Gillespie, Karan Goel, Noah Goodman, Shelby Grossman, Neel Guha, Tatsunori
  Hashimoto, Peter Henderson, John Hewitt, Daniel~E. Ho, Jenny Hong, Kyle Hsu,
  Jing Huang, Thomas Icard, Saahil Jain, Dan Jurafsky, Pratyusha Kalluri,
  Siddharth Karamcheti, Geoff Keeling, Fereshte Khani, Omar Khattab, Pang~Wei
  Koh, Mark Krass, Ranjay Krishna, Rohith Kuditipudi, Ananya Kumar, Faisal
  Ladhak, Mina Lee, Tony Lee, Jure Leskovec, Isabelle Levent, Xiang~Lisa Li,
  Xuechen Li, Tengyu Ma, Ali Malik, Christopher~D. Manning, Suvir Mirchandani,
  Eric Mitchell, Zanele Munyikwa, Suraj Nair, Avanika Narayan, Deepak
  Narayanan, Ben Newman, Allen Nie, Juan~Carlos Niebles, Hamed Nilforoshan,
  Julian Nyarko, Giray Ogut, Laurel Orr, Isabel Papadimitriou, Joon~Sung Park,
  Chris Piech, Eva Portelance, Christopher Potts, Aditi Raghunathan, Rob Reich,
  Hongyu Ren, Frieda Rong, Yusuf Roohani, Camilo Ruiz, Jack Ryan, Christopher
  Ré, Dorsa Sadigh, Shiori Sagawa, Keshav Santhanam, Andy Shih, Krishnan
  Srinivasan, Alex Tamkin, Rohan Taori, Armin~W. Thomas, Florian Tramèr,
  Rose~E. Wang, William Wang, Bohan Wu, Jiajun Wu, Yuhuai Wu, Sang~Michael Xie,
  Michihiro Yasunaga, Jiaxuan You, Matei Zaharia, Michael Zhang, Tianyi Zhang,
  Xikun Zhang, Yuhui Zhang, Lucia Zheng, Kaitlyn Zhou, and Percy Liang.
\newblock On the opportunities and risks of foundation models, 2022.

\bibitem{Marcus_2021-kr}
Gary Marcus.
\newblock Has {AI} found a new foundation?
\newblock \url{https://thegradient.pub/has-ai-found-a-new-foundation/}, 2021.
\newblock Accessed: 2023-4-17.

\bibitem{Ridnik_2021}
Tal Ridnik, Emanuel Ben-Baruch, Asaf Noy, and Lihi Zelnik.
\newblock Imagenet-21k pretraining for the masses.
\newblock In J.~Vanschoren and S.~Yeung, editors, {\em Proceedings of the
  Neural Information Processing Systems Track on Datasets and Benchmarks},
  volume~1. Curran, 2021.

\bibitem{wang2022omnivlone}
Junke Wang, Dongdong Chen, Zuxuan Wu, Chong Luo, Luowei Zhou, Yucheng Zhao,
  Yujia Xie, Ce~Liu, Yu-Gang Jiang, and Lu~Yuan.
\newblock Omnivl:one foundation model for image-language and video-language
  tasks, 2022.

\bibitem{yu2022coca}
Jiahui Yu, Zirui Wang, Vijay Vasudevan, Legg Yeung, Mojtaba Seyedhosseini, and
  Yonghui Wu.
\newblock Coca: Contrastive captioners are image-text foundation models, 2022.

\bibitem{yuan2021florence}
Lu~Yuan, Dongdong Chen, Yi-Ling Chen, Noel Codella, Xiyang Dai, Jianfeng Gao,
  Houdong Hu, Xuedong Huang, Boxin Li, Chunyuan Li, Ce~Liu, Mengchen Liu,
  Zicheng Liu, Yumao Lu, Yu~Shi, Lijuan Wang, Jianfeng Wang, Bin Xiao, Zhen
  Xiao, Jianwei Yang, Michael Zeng, Luowei Zhou, and Pengchuan Zhang.
\newblock Florence: A new foundation model for computer vision, 2021.

\bibitem{xie2021selfsupervised}
Zhenda Xie, Yutong Lin, Zhuliang Yao, Zheng Zhang, Qi~Dai, Yue Cao, and Han Hu.
\newblock Self-supervised learning with swin transformers, 2021.

\bibitem{Moor2023-fa}
Michael Moor, Oishi Banerjee, Zahra Shakeri~Hossein Abad, Harlan~M Krumholz,
  Jure Leskovec, Eric~J Topol, and Pranav Rajpurkar.
\newblock Foundation models for generalist medical artificial intelligence.
\newblock {\em Nature}, 616(7956):259--265, April 2023.

\bibitem{bucher2019zeroshot}
Maxime Bucher, Tuan-Hung Vu, Matthieu Cord, and Patrick Pérez.
\newblock Zero-shot semantic segmentation, 2019.

\bibitem{9665941}
Björn Michele, Alexandre Boulch, Gilles Puy, Maxime Bucher, and Renaud Marlet.
\newblock Generative zero-shot learning for semantic segmentation of 3d point
  clouds.
\newblock In {\em 2021 International Conference on 3D Vision (3DV)}, pages
  992--1002, 2021.

\bibitem{Luddecke_2022_CVPR}
Timo L\"uddecke and Alexander Ecker.
\newblock Image segmentation using text and image prompts.
\newblock In {\em Proceedings of the IEEE/CVF Conference on Computer Vision and
  Pattern Recognition (CVPR)}, pages 7086--7096, June 2022.

\bibitem{pmlr-v139-radford21a}
Alec Radford, Jong~Wook Kim, Chris Hallacy, Aditya Ramesh, Gabriel Goh,
  Sandhini Agarwal, Girish Sastry, Amanda Askell, Pamela Mishkin, Jack Clark,
  Gretchen Krueger, and Ilya Sutskever.
\newblock Learning transferable visual models from natural language
  supervision.
\newblock In Marina Meila and Tong Zhang, editors, {\em Proceedings of the 38th
  International Conference on Machine Learning}, volume 139 of {\em Proceedings
  of Machine Learning Research}, pages 8748--8763. PMLR, 18--24 Jul 2021.

\bibitem{he2023computervision}
Sheng He, Rina Bao, Jingpeng Li, Jeffrey Stout, Atle Bjornerud, P.~Ellen Grant,
  and Yangming Ou.
\newblock Computer-vision benchmark segment-anything model (sam) in medical
  images: Accuracy in 12 datasets, 2023.

\bibitem{zhang2023customized}
Kaidong Zhang and Dong Liu.
\newblock Customized segment anything model for medical image segmentation,
  2023.

\bibitem{hu2021lora}
Edward~J. Hu, Yelong Shen, Phillip Wallis, Zeyuan Allen-Zhu, Yuanzhi Li, Shean
  Wang, Lu~Wang, and Weizhu Chen.
\newblock Lora: Low-rank adaptation of large language models, 2021.

\bibitem{ma2023segment}
Jun Ma and Bo~Wang.
\newblock Segment anything in medical images, 2023.

\bibitem{Hosny2018-pd}
Ahmed Hosny, Chintan Parmar, John Quackenbush, Lawrence~H Schwartz, and Hugo J
  W~L Aerts.
\newblock Artificial intelligence in radiology.
\newblock {\em Nat. Rev. Cancer}, 18(8):500--510, August 2018.

\bibitem{Willemink2020-rs}
Martin~J Willemink, Wojciech~A Koszek, Cailin Hardell, Jie Wu, Dominik
  Fleischmann, Hugh Harvey, Les~R Folio, Ronald~M Summers, Daniel~L Rubin, and
  Matthew~P Lungren.
\newblock Preparing medical imaging data for machine learning.
\newblock {\em Radiology}, 295(1):4--15, April 2020.

\bibitem{Federer2015-hu}
Lisa~M Federer, Ya-Ling Lu, Douglas~J Joubert, Judith Welsh, and Barbara
  Brandys.
\newblock Biomedical data sharing and reuse: Attitudes and practices of
  clinical and scientific research staff.
\newblock {\em PLoS One}, 10(6):e0129506, June 2015.

\bibitem{Sunoqrot2022-zm}
Mohammed R~S Sunoqrot, Anindo Saha, Matin Hosseinzadeh, Mattijs Elschot, and
  Henkjan Huisman.
\newblock Artificial intelligence for prostate {MRI}: open datasets, available
  applications, and grand challenges.
\newblock {\em Eur Radiol Exp}, 6(1):35, August 2022.

\bibitem{Cushnan2021-zx}
Dominic Cushnan, Rosalind Berka, Ottavia Bertolli, Peter Williams, Daniel
  Schofield, Indra Joshi, Alberto Favaro, Mark Halling-Brown, Gergely Imreh,
  Emily Jefferson, Neil~J Sebire, Gerry Reilly, Jonathan C~L Rodrigues, Graham
  Robinson, Susan Copley, Rizwan Malik, Claire Bloomfield, Fergus Gleeson,
  Moira Crotty, Erika Denton, Jeanette Dickson, Gary Leeming, Hayley~E
  Hardwick, Kenneth Baillie, Peter~Jm Openshaw, Malcolm~G Semple, Caroline
  Rubin, Andy Howlett, Andrea~G Rockall, Ayub Bhayat, Daniel Fascia, Cathie
  Sudlow, {NCCID Collaborative}, and Joseph Jacob.
\newblock Towards nationally curated data archives for clinical radiology image
  analysis at scale: Learnings from national data collection in response to a
  pandemic.
\newblock {\em Digit Health}, 7:20552076211048654, November 2021.

\bibitem{Menze2015-kv}
Bjoern~H Menze, Andras Jakab, Stefan Bauer, Jayashree Kalpathy-Cramer, Keyvan
  Farahani, Justin Kirby, Yuliya Burren, Nicole Porz, Johannes Slotboom, Roland
  Wiest, Levente Lanczi, Elizabeth Gerstner, Marc-Andr{\'e} Weber, Tal Arbel,
  Brian~B Avants, Nicholas Ayache, Patricia Buendia, D~Louis Collins, Nicolas
  Cordier, Jason~J Corso, Antonio Criminisi, Tilak Das, Herv{\'e} Delingette,
  {\c C}a{\u g}atay Demiralp, Christopher~R Durst, Michel Dojat, Senan Doyle,
  Joana Festa, Florence Forbes, Ezequiel Geremia, Ben Glocker, Polina Golland,
  Xiaotao Guo, Andac Hamamci, Khan~M Iftekharuddin, Raj Jena, Nigel~M John,
  Ender Konukoglu, Danial Lashkari, Jos{\'e}~Antoni{\'o} Mariz, Raphael Meier,
  S{\'e}rgio Pereira, Doina Precup, Stephen~J Price, Tammy~Riklin Raviv, Syed
  M~S Reza, Michael Ryan, Duygu Sarikaya, Lawrence Schwartz, Hoo-Chang Shin,
  Jamie Shotton, Carlos~A Silva, Nuno Sousa, Nagesh~K Subbanna, Gabor Szekely,
  Thomas~J Taylor, Owen~M Thomas, Nicholas~J Tustison, Gozde Unal, Flor
  Vasseur, Max Wintermark, Dong~Hye Ye, Liang Zhao, Binsheng Zhao, Darko Zikic,
  Marcel Prastawa, Mauricio Reyes, and Koen Van~Leemput.
\newblock The multimodal brain tumor image segmentation benchmark ({BRATS}).
\newblock {\em IEEE Trans. Med. Imaging}, 34(10):1993--2024, October 2015.

\bibitem{Saha2022-ki}
Anindo Saha, Jasper~Jonathan Twilt, Joeran~Sander Bosma, Bram van Ginneken,
  Derya Yakar, Mattijs Elschot, Jeroen Veltman, Jurgen F{\"u}tterer, Maarten
  de~Rooij, and Henkjan Huisman.
\newblock The {PI-CAI} challenge: Public training and development dataset, May
  2022.

\bibitem{Yang2023-hy}
Jiancheng Yang, Rui Shi, Donglai Wei, Zequan Liu, Lin Zhao, Bilian Ke,
  Hanspeter Pfister, and Bingbing Ni.
\newblock {MedMNIST} v2 - a large-scale lightweight benchmark for {2D} and {3D}
  biomedical image classification.
\newblock {\em Sci Data}, 10(1):41, January 2023.

\bibitem{Liu2021-fz}
Xiangbin Liu, Liping Song, Shuai Liu, and Yudong Zhang.
\newblock A review of {Deep-Learning-Based} medical image segmentation methods.
\newblock {\em Sustain. Sci. Pract. Policy}, 13(3):1224, January 2021.

\bibitem{Antonelli2022-yk}
Michela Antonelli, Annika Reinke, Spyridon Bakas, Keyvan Farahani, Annette
  Kopp-Schneider, Bennett~A Landman, Geert Litjens, Bjoern Menze, Olaf
  Ronneberger, Ronald~M Summers, Bram van Ginneken, Michel Bilello, Patrick
  Bilic, Patrick~F Christ, Richard K~G Do, Marc~J Gollub, Stephan~H Heckers,
  Henkjan Huisman, William~R Jarnagin, Maureen~K McHugo, Sandy Napel, Jennifer
  S~Golia Pernicka, Kawal Rhode, Catalina Tobon-Gomez, Eugene Vorontsov,
  James~A Meakin, Sebastien Ourselin, Manuel Wiesenfarth, Pablo Arbel{\'a}ez,
  Byeonguk Bae, Sihong Chen, Laura Daza, Jianjiang Feng, Baochun He, Fabian
  Isensee, Yuanfeng Ji, Fucang Jia, Ildoo Kim, Klaus Maier-Hein, Dorit Merhof,
  Akshay Pai, Beomhee Park, Mathias Perslev, Ramin Rezaiifar, Oliver Rippel,
  Ignacio Sarasua, Wei Shen, Jaemin Son, Christian Wachinger, Liansheng Wang,
  Yan Wang, Yingda Xia, Daguang Xu, Zhanwei Xu, Yefeng Zheng, Amber~L Simpson,
  Lena Maier-Hein, and M~Jorge Cardoso.
\newblock The medical segmentation decathlon.
\newblock {\em Nat. Commun.}, 13(1):4128, July 2022.

\bibitem{Armato2018-jv}
Samuel~G Armato, 3rd, Henkjan Huisman, Karen Drukker, Lubomir Hadjiiski,
  Justin~S Kirby, Nicholas Petrick, George Redmond, Maryellen~L Giger, Kenny
  Cha, Artem Mamonov, Jayashree Kalpathy-Cramer, and Keyvan Farahani.
\newblock {PROSTATEx} challenges for computerized classification of prostate
  lesions from multiparametric magnetic resonance images.
\newblock {\em J Med Imaging (Bellingham)}, 5(4):044501, October 2018.

\bibitem{Zabihollahy2019-ep}
Fatemeh Zabihollahy, Nicola Schieda, Satheesh Krishna~Jeyaraj, and Eranga
  Ukwatta.
\newblock Automated segmentation of prostate zonal anatomy on t2-weighted
  ({T2W}) and apparent diffusion coefficient ({ADC}) map {MR} images using
  {U-Nets}.
\newblock {\em Med. Phys.}, 46(7):3078--3090, July 2019.

\bibitem{deng2023segment}
Ruining Deng, Can Cui, Quan Liu, Tianyuan Yao, Lucas~W. Remedios, Shunxing Bao,
  Bennett~A. Landman, Lee~E. Wheless, Lori~A. Coburn, Keith~T. Wilson, Yaohong
  Wang, Shilin Zhao, Agnes~B. Fogo, Haichun Yang, Yucheng Tang, and Yuankai
  Huo.
\newblock Segment anything model (sam) for digital pathology: Assess zero-shot
  segmentation on whole slide imaging, 2023.

\end{thebibliography}

\end{document}